# CdZnTe:Cl crystals for X-ray computer tomography detectors

O. A. Matveev (a), A. I. Terent'ev[1]) (a), V. P. Karpenko (a), N. K. Zelenina (a), A. Fauler (b), M. Federle (b), and K. W. Benz (b)

(a) Ioffe Physico-Technical Institute, Russian Academy of Sciences, Politecknicheskaya 26, St. Petersburg 194021, Russia

(b) Kristallographisches Institut, Albert-Ludwigs-Universitat Freiburg, Hebelstrasse 25, D-79104 Freiburg, Germany

Processes of growth of semi-insulating $Cd_{1-x}Zn_xTe$: Cl crystals ($x$ = 0.0002 and 0.1) of n-type conductivity are investigated. From the grown crystals detectors for X-ray computer tomography with small value of photocurrent memory (afterglow) (0.1-0.3%) are obtained.

The interest in examination of CdTe and CdZnTe crystals for tomographic detectors is clear. It is well-known that CdTe:Cl is a promising material for X-ray computer tomography (CT) detectors [1-3]. However, when the first CdTe tomographic detectors were created a material with a low concentration of traps $(N_{tr} < 10^{14}$ cm$^{-3})$ was found to be necessary for obtaining of a small photocurrent memory (afterglow). Such CdTe crystals are not obtained till now. It was shown [4] that the negative role of traps can be reduced by introduction of an additional concentration $(N_r)$ of recombination ceners $V_{CD}^{-2} \sim (E_V + 0.9$ eV) so that $N_{tr} \sim N_r \sim 10^{15}$ cm$^{-3}$ by means of post growth annealing of a crystal. The detectors built around such crystals satisfied the requirements of medical X-ray CT [5]. The wide use of CdTe:Cl for CT detectors was mainly restrained by the low reproducibility of obtaining semi-insulating crystals with the required semiconductor performances. It is caused by the fundamental properties of the material. CdTe is rather predisposed to different structural imperfections mainly due to the "weakness" of the cadmium sublattice (small energies of $V_{Cd}$ and $Cd_I$ formation).

Introduction of Zn stabilizes the cadmium sublattice of CdTe crystals [6, 7] and essentially improves the structural characteristics of the grown crystals. It is known [8, 9] that Zn also influences the structure of energy levels in the forbidden gap, and consequently the performance of detector material.

In this paper, the role of Zn in obtaining semi-insulating CdZnT:Cl crystals for X-ray CT detectors was investigated.

The growth of CdTe and CdZnTe crystals was carried out by a method of temperature gradient freezing in a horizontal furnace under controlled Cd vapor pressure $(P_{Cd})$. The solidification of the material occurs in a furnace with no moving parts (the temperature is controlled using programmed power reduction). This process results in a slow motion of the slightly concave crystallization front into the melt. Precision of the temperature control was 0.25 °C.

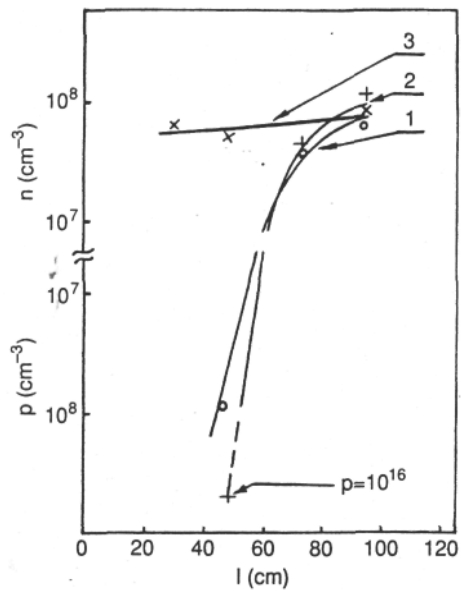

Fig. 1. Concentration of free carriers on the length of the ingot. Curves 1, 2 correspond to temperature gradients of 2 °C/cm and curve 3 is for 1.3°C/cm; Zn content: (1) $x$ = 0.1, (2) 0.05, (3) 0.0002

The melt at the end of the boat is overheated by an amount given by the product of the temperature gradient and the boat length in comparison with the melting point. High overheating of the melt results in the formation of pores in the ingot as well as in change of both melt and crystal compositions and, therefore in the electrical performances of a crystal [10]. With the account of the boat length grad $T$ < 2 °C/cm was chosen which allowed work with a stable melt. Decrease of grad $T$ allows any difference of melt compositions between the beginning and the end of the boat to be reduced as well as convective streams in the fluid [10] to be lowered. Figure 1 shows the variation of concentration of free carriers along the ingots grown by us with grad $T$ = 2 and 1.3 °C/cm. In the first case (curves 1, 2) we can see a profound change of carrier concentration and an inversion of the conductance type from the beginning to the end of the ingot. In the second case (curve 3) an ingot with almost homogeneous electron concentration along its length is obtained. The best results at (grad $T$) — 1.3 °C/cm were established to be .obtained at a growth rate 0.7 cm/h.

It is desirable to carry out crystal growth at conditions where compositions of the melt and the crystal are identical (S = L) and inclusions of the second phase are not formed. However, it is very difficult to reproduce these requirements precisely. Therefore, the crystals were grown with a small enrichment of the melt composition by cadmium. Thus, due to the excess cadmium formed at the crystallization front having time to evaporate, the second phase was not formed [11]. Keeping the chosen requirements the grown crystals had good homogeneity of electrical performances both along the ingot and in its cross section.

CdZnTe:Cl having been grown during the ingot was annealed [12] to remove the grown-in stresses arising from the axial temperature gradient and the nonplanar shape of the crystallization front. The annealing was done to yield a composition of the crystal corresponding to the best conditions for self-compensation. Post-growth annealing of the ingot was carried out in the same ampoule without its decompression by means of the same system of furnaces. The homogeneous composition of CdZnTe:Cl in the ingot and removal of stresses in its volume is achieved by gradientless thermal homogenizing annealing. The best results have, been received by 6h annealing immediately after growth.

Further annealing of the ingot consists in its programmed cooling under controlled $P_{Cd}$ both at high-temperature ($T_{ann}$ = 1070-800 °C) and low-temperature ($T_{ann}$, = 800 -400 °C) stages [12]. At the high-temperature stage the crystal cooling rate was chosen small as well as $P_{Cd}$ reducing rate in order to achieve a concentration agreement between the basic donor ($Cl_{Te+}$) and acceptor ($V_{Cd}^{-2}$) defects throughout the ingot. Chlorine solubility dependence on temperature was taken into account [12]. Semi-insulating material ($\rho \ll 10^9$ Ω cm) is obtained in the whole interval of doping by chlorine (2 x $10^{18}$ to 2 x $10^{19}$ $cm^{-3}$) when the cooling rate is equal to 2 °C/h [12].

At the low-temperature stage of annealing a high rate of cooling (50-80 °C/h) was chosen so that the agreement between the charged defect concentrations achieved at a high temperature stage had no time to be disturbed. But this cooling rate allowed the charged defects to be combined into neutral complexes via diffusion.

It is very difficult to receive CdTe:Cl semi-insulating crystals of n-type conductance with the homogeneous ingot electron concentration. These crystals can be obtained only if $P_{Cd}$ corresponds to minimum $n$ in dependence of $n$ on $P_{Cd}$ [13]. But it is difficult to achieve such conditions because of the very *sharp* character of this minimum. A similar minimum of free carrier concentration being expected on CdZnTe:Cl crystals as well, therefore, earlier results [13] were used to obtain these crystals.

ZnTe crystals are known to have p-type conductance only [14]. Therefore, the variation of conductance from n type to p one having increased Zn content should be expected in the solid $Cd_{1-x}Zn_xTe$ solution. However, at $x$ up to 0.1 we did not observe it. At identical *P-T* annealing conditions of ingots with $x$ = 0.0002 and $x$ = 0.1 we have obtained crystals with equal concentrations of $n \sim 10^7$ $cm^{-3}$. Under these conditions semi-insulating crystals only of *p*-type of conductance were obtained when CdTe:Cl was grown. These results are the evidence that introduction of zinc promotes the formation of deep centers in $Cd_{1-x}Zn_xTe$:Cl in addition to the mechanism of self-compensation.

Addition of Zn to CdTe results in chemical bond strengthening in the lattice [15] which should in turn cause the vacancy concentration in the metal sublattice to stabilize. Really, the solubility region of metal vacancies is increased by more than an order of magnitude in ZnTe [16]. $Cd_{1-x}Zn_xTe$ crystals have been recently established [17] to contain mainly neutral metal divacancies.

$\mu\tau_e$ and, $\mu\tau_h$ of $Cd_{1-x}Zn_xTe$:Cl crystals are about two and ten times smaller, respectively, than those of CdTe:Cl crystals [12] them (it is very important!) varying only slightly along the ingot. These $\mu\tau_e$ and, $\mu\tau_h$ values are equal for samples with Zn contents of $x$ = 0.0002 and $x$ = 0.1. Apparently, this result is also indicative of the formation of an additional deep recombination center in the presence of zinc in the crystal.

Spectra of the extrinsic photoconductivity of the grown $Cd_{1-x}Zn_xTe$:Cl crystals *($x$ = 0.0002 and 0.1)* and CdTe:Cl crystals confirm all the above-mentioned facts. Figure 2 shows the spectra of crystal CdTe:Cl extrinsic photoconductivity for spectrometer detectors (curve 1), tomographic detectors (curve 2) and crystals $Cd_{1-x}Zn_xTe$:Cl for $x$ = 0.0002 and 0.1 (curves 3 and 4, respectively).

As expected, with the use of postgrowth annealing [4] it became possible to increase the concentration of deep recombination centers ($E_v + 0.9$) eV (Fig. 2, curve 2 in comparison to curve 1). It allowed after-glowing in tomographic M-S-M detectors made from such material to be tenfold (up to several percents) reduced.

The deep center concentration rise with Zn contents increase is seen (Fig. 2, curves 3 and 4) in crystals of $Cd_{1-x}Zn_xTe:Cl$. Our study of characteristics of the M-S-M detectors made from these crystals with injecting contacts in a regime of double injection [5] has shown that even at low Zn concentration ($x = 0.0002$) afterglow decreases up to 0.1%-0.3%, linearity of signal dependence on radiation stream intensity approaches to unity, and sensitivity practically does not change as compared to specially annealed crystals (Fig. 2, curve 2). Highly efficient, tomographic M-S-M detectors have been built around grown $Cd_{1-x}Zn_xTe:Cl$ crystals. The noise was very low (~5 x $10^{-12}$ A) and the response to X-ray radiation was quick (~$10^{-4}$ s). Compared to CdTe:Cl crystals the photocur-rent memory (so called afterglow) was significantly less (< 0.3%). The signal dependence on stream intensity of radiation in the range of ~$10^3$ times was linear. These detectors being possible to work at room temperature as well as being noteworthy for their small sizes (2 x 2 x 10 mm$^3$) may be used for creation of detecting blocks for X-ray medical and technical tomography.

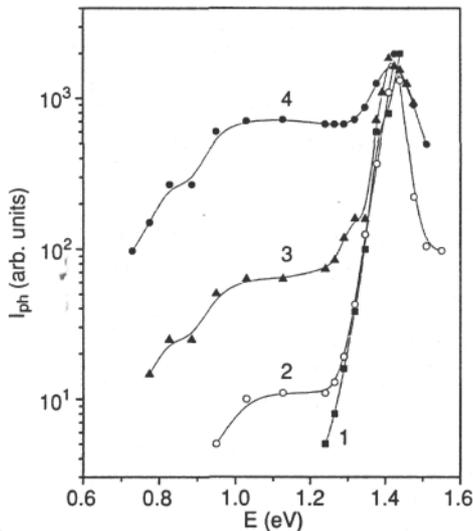

Fig. 2. Spectra of extrinsic photoconductivity of CdTe:Cl crystals (curve 1,2) and $Cd_{1-x}Zn_xTe:Cl$ (curves 3, 4); (1) for spectrometer detectors,
(2) for tomographic detectors) (curves 3, 4),
(3) $x = 0.0002$ and (4) 0.01

*Acknowledgement* The studies were supported by INTAS grant No. 99-1456.

**References**

1] E. N. ARKAD'EVA et al., Zh. Tekh. Rz. 51,1933 (1981).
2] P. A. GLASOV et al., IEEE Trans. Nucl. Sci. 28, 563 (1981).
3] S. RICQ, F. GLASER, and M. GARCIN, Nucl Instrum. Methods Phys. Res. A 458, 534 (2001).
4] E. N. ARKAD'EVA et al., Sov. Phys. - Tech. Phys. 30, 535 (1985).
5] N. K. ZELENINA et al., Nucl. Instrum. Methods Phys. Res A 283, 36 (1989).
[6] K. GUERGOURY, R. TRIBOULET, A. THOMSON-CARLY, and Y. MARFAING, J. Cryst. Growth 86, 61 (1988).
[7] S. SEN et al., J. Cryst. Growth 86, 111 (1988).
[8] Cs. SZELES et al., Nucl. Instrum. Methods Phys. Res. A 380,148 (1996).
[9] A. ZERRAI et al., J. Cryst. Growth 197, 646 (1999).
[10] O. A.MATVEEV and A. I. TERENT'EV, Semiconductors 29,191 (1995).
[11] D. DE NOBEL, Philips Res. Rep. 14, 361 (1959).
[12] O. A. MATVEEV and A. I. TERENT'EV, Semiconductors 34,1264 (2000).
[13] F. F. MOREHEAD and G. MANDEL, Appl. Phys. Lett. 5, 53 (1964).
[14] Physical-Chemical Properties of Semiconductor Substances (Reference Book), Izd. Nauka, Moscow 1979 (p. 118).
[15] O. A. MATVEEV and A. I. TERENT'EV, Semiconductors 32,144 (1998).
[16]S. B. QADRI, E. F. SKELTON, and A. W. WEBB, Appl. Phys. Lett. 46, 257 (1985).
[17] G. TESSARO and P. MASCHER, J. Cryst. Growth 179, 581 (1999).